\documentclass[10pt,conference,a4paper]{IEEEtran}
\usepackage[T1]{fontenc}
\usepackage{amsmath}
\usepackage[cmintegrals]{newtxmath}
\usepackage{url}
\usepackage{soul,color,graphicx,float,epstopdf}
\usepackage{tablefootnote,textcomp,gensymb}
\usepackage{eurosym,booktabs,multirow}
\usepackage[caption=false]{subfig}
\usepackage{amsfonts} 
\usepackage{algorithmic}
\usepackage{array}
\usepackage{stfloats}
\usepackage{float}
\usepackage{mathtools}
\usepackage{graphicx}
\usepackage{pdfpages}
\usepackage{subfig} 
\usepackage{balance}
\usepackage{xcolor}
\usepackage{comment}
\usepackage[normalem]{ulem}
\usepackage{dsfont}
\usepackage[linesnumbered, ruled]{algorithm2e}
\usepackage{enumitem}
 
\usepackage{tikz}
\usetikzlibrary{arrows,shapes,positioning,shadows,trees}

\tikzset{
  basic/.style  = {draw, text width=4cm, drop shadow, font=\sffamily, rectangle},
  root/.style   = {basic, rounded corners=2pt, thin, align=center,
                   fill=red!20},
  level 2/.style = {basic, rounded corners=6pt, thin,align=center, fill=blue!20,
                   text width=9em},
  level 3/.style = {basic, thin, align=left, fill=pink!20, text width=8em}
}

\definecolor{Orange}{rgb}{1,0.5,0}
\newcommand{\eqr}[1]{$#1$}

\newcommand*{\rom}[1]{\expandafter\@slowromancap\romannumeral #1@}

\newcommand{\PZ}[1]{\textcolor{black}{#1}}

\begin{document}

\title{Federated Meta-Learning for Traffic Steering in O-RAN\\
}

\author{\IEEEauthorblockN{Hakan Erdol\IEEEauthorrefmark{1},
Xiaoyang Wang\IEEEauthorrefmark{1},
Peizheng Li\IEEEauthorrefmark{1},
Jonathan D. Thomas\IEEEauthorrefmark{1},
Robert Piechocki\IEEEauthorrefmark{1}, \\
George Oikonomou\IEEEauthorrefmark{1},
Rui Inacio\IEEEauthorrefmark{2},
Abdelrahim Ahmad\IEEEauthorrefmark{2},
Keith Briggs\IEEEauthorrefmark{3},
Shipra Kapoor \IEEEauthorrefmark{3}
}\\ 
\IEEEauthorblockA{\IEEEauthorrefmark{1} University of Bristol, UK;
\IEEEauthorrefmark{2} Vilicom UK Ltd. ; \IEEEauthorrefmark{3} Applied  Research, BT, UK \\
Email: {\{hakan.erdol, xiaoyang.wang, peizheng.li, jonathan.david.thomas, r.j.piechocki, g.oikonomou \}@bristol.ac.uk}\\
{\{Rui.Inacio, Abdelrahim.Ahmad\}@vilicom.com};
{\{keith.briggs, shipra.kapoor, \}@bt.com}
}}

\maketitle

\begin{abstract}
The vision of 5G lies in providing high data rates, low latency (for the aim of near-real-time applications), significantly increased base station capacity, and near-perfect quality of service (QoS) for users, compared to LTE networks. In order to provide such services, 5G systems will support various combinations of access technologies such as LTE, NR, NR-U and Wi-Fi. Each radio access technology (RAT) provides different types of access, and these should be allocated and managed optimally among the users. Besides resource management, 5G systems will also support a dual connectivity service. The orchestration of the network therefore becomes a more difficult problem for system managers with respect to legacy access technologies. In this paper, we propose an algorithm for RAT allocation based on federated meta-learning (FML), which enables RAN intelligent controllers (RICs) to adapt more quickly to dynamically changing environments. We have designed a simulation environment which contains LTE and 5G NR service technologies. In the simulation, our objective is to fulfil UE demands within the deadline of transmission to provide higher QoS values. We compared our proposed algorithm with a single RL agent, the Reptile algorithm and a rule-based heuristic method. Simulation results show that the proposed FML method achieves higher caching rates at first deployment round 21\% and 12\% respectively. Moreover, proposed approach adapts to new tasks and environments most quickly amongst the compared methods. 
\end{abstract}

\begin{IEEEkeywords}
Federated  Meta Learning, Reinforcement learning , Traffic steering, 5G, O-RAN, Resource management.
\end{IEEEkeywords}

\section{Introduction}
The next generation of networks need to support a wide range of demanding applications such as connected autonomous vehicles, 4k video streaming, and ultra-low-power Internet-of-things networks. This variety of services prevents service providers having one simple solution for all users. In order to provide such services, more intelligent and agile solutions are required for mobile network users. To this end, open radio access network (O-RAN) enables radio intelligent controllers (RIC) units to provide a resource management scheme to tackle different problems within the same structure; this is called traffic steering. 

Traffic steering is one of the many use cases of O-RAN. There are numerous types of resources needing to be allocated in 5G systems. In addition to cell allocation and band allocation, 5G systems support different combinations of access technologies namely, NR (New radio, licensed band), NR-U (unlicensed band), LTE and Wi-Fi \cite{ORAN_arc}. Moreover, 5G systems will provide dual connectivity to UEs e.g., while mobile broadband (MBB) service is provided by one cell, voice and low data rate required applications can be provided by other cells. 
The RIC requires controlling many combinations to provide high QoS to the users \cite{ORAN_TS_AI}. 

Traffic steering is a hard problem which becomes more difficult the larger the number of participants. Thus, heuristic methods often fail to fulfil such high QoS demand. Therefore, machine learning-based algorithms are a more promising way to solve this problem. Machine learning (ML)-based resource allocation for wireless networks have been studied for over a decade \cite{RL_res1}. Yet new problems are still emerging, with new demands and new radio access technologies (RAT) such as effectively managing 5G new radio (NR) features. ML-based algorithms for resource allocation problems often use reinforcement learning (RL)-based methods \cite{ORAN_TS_AI}. These methods use a simulated environment to generate data to train RL agents by using these data. While there are number of RL formulations, we used a deep reinforcement learning (DRL) method for decision making in simulated environment.

RL agents that have stochastic environments require an adaptation phase to achieve optimal rewards. In order to increase the convergence speed of model adaptation to the environment, a collaborative learning algorithm, namely federated meta-learning (FML) \cite{FedMeta2022} is proposed to orchestrate the network in the O-RAN environment.

There are several reasons to choose the FL framework for the O-RAN ecosystem. One of the main reasons is to generalise across the distribution of environments for RL agents \cite{FRL_converge}. Cellular networks contain highly dynamic and unique environments. Even well-trained RL agents may fail to adapt to the environment after deployment. If RIC management cannot deal with a quickly changing environment, it can cause significant QoE issues for users. Another reason is some areas may have different priorities than others in aspects of service types; for example, some applications may demand higher throughput, and some may need lower latency while communicating. 
We defined several QoS metrics such as throughput and latency to generalize the problem within the FML framework.

\PZ{Our motivation for using meta-learning approach for DRL algorithm is to enhance the adaptation process of RL agents. The reason for focusing on the adaptation process is that wireless communication in RAN is highly dynamic. Moreover, service demands are application-dependent (as mentioned before) and, obtaining optimal solution faster plays a crucial role in intelligent resource management applications. Hence, we aim to train a DQN model that enables RL agent adapt to a new task i.e., latency, throughput, or caching rate, can be quick.}

We propose the form of FML which uses the reptile algorithm \cite{nichol2018Reptile} for meta-learning.

The major contributions of this paper are as follows:

\begin{itemize}
    \item A RAT allocation environment which enables RL agents to train their DQN models for steering traffic between RATs to provide service to vehicles.
    \item Various QoS performance metrics are measured in the environment. These QoS metrics are defined as unique tasks for the meta learning algorithm.
    \item A federated meta-learning framework is designed for higher convergence speeds to unseen tasks and environments. We distributed learning algorithms in the framework and analysed the results.
    \item We evaluated how the rule-based approach and learning-based approaches are performing in our RAT allocation environment. Results show that the proposed FML algorithm performs the best among other approaches.
    \item To the best of our knowledge, there is no paper that simulates a distributed setup for traffic steering which is supported by O-RAN architecture. 
\end{itemize}

The rest of the paper is organized as follows: In Section II, related works are discussed. In section III, the system model is described and distinguished from the contributions of this work. Section IV describes the Markov decision process model for the simulation
model and parameters. Section V describes the proposed FL framework to solve a simplified traffic steering problem while simulation results are presented in Section
VI. We evaluate our findings and outline courses for future
work in Section VII.

\section{Background}

Federated learning (FL) \cite{FL_original} paradigm aims to perform training directly on edge devices with their own local data and employs a communication efficient model management scheme. While collected data is kept local, communication between the global server and edge devices contains only model gradients \cite{FL_original}. Therefore, FL is a communication and computation-efficient way of distributed learning algorithms. The server merges these updates to form a better more generic model for all environments. Model updates at local RL agents represent information about local environments. Thus, the global model collects representative information about the deployed environment without getting any state data. This feature prevents both constant data flow from deployed units and keeps private data local such as user equipment's (UE) locations, routines, data usage, etc. After aggregating model updates at the server, the server forms a global model as a generic model for all agents.

\PZ{ Federated reinforcement learning (FRL) enables RL agents to generalise environment by using collaborative scheme. Liu et al. \cite{FRL_converge} used FL framework to extend RL agents' experience so that they can effectively use prior knowledge. Their objective is to improve adaptation speed to a new environment by using this knowledge. Our proposed method uses various environments for the same motivation as well. However, in addition to FL framework, we also employ meta-learning to enable RL agents to adapt faster to new QoS tasks as well as environments.}

The FML algorithm is used to increase converging speeds to unseen tasks \cite{FedMeta2020}. This feature enables RL agents or any other DNN-based learning algorithms to adapt to new tasks faster. \PZ{Yue et al. \cite{FedMeta2022} used FML approach to maximize the theoretical lower bound of global loss reduction in each round to accelerate the convergence. Unlike our approach, they added user selection mechanism according to contributions of local models.} 

\PZ{Besides RL-based methods, there is also recurrent neural network (RNN) based FML methods \cite{FML2022_UoB}. Zhang et al. used different cities as meta-tasks to train an RNN model to predict network traffic. While they used each city as task, we used each QoS metric in the network as a meta-task in this paper. In both way, tasks are not completely independent. Network traffic for cities usually represents a seasonal behavior that changes accord the  time of the day. Therefore, traffic demand changes at similar times but in different volumes. In our paper, we used different QoS metrics as tasks and they are indirectly dependent as well e.g., higher throughput will lead to transmit data in less time and it will decrease the latency.}

There are many ML-based resource allocation papers in the literature \cite{5G_survey}. However, there are only a few studies that use ML to provide a solution to traffic steering use case since it is relatively new problem. \PZ{Adamczyk et al. \cite{RL_TS_2021} used an artificial neural network trained with the SARSA algorithm for traffic control in RANs. For traffic steering use case, they allocated resources among the users according to base station load. They defined different user profiles according to their data rate demands. In our paper, we define these demands as meta-tasks. Thus, after the deployment of an RL agent,  it adapts to a new demand profile more rapidly.}
\section{System Model}
\begin{figure}[t]
\centering
\includegraphics[width=0.8\columnwidth]{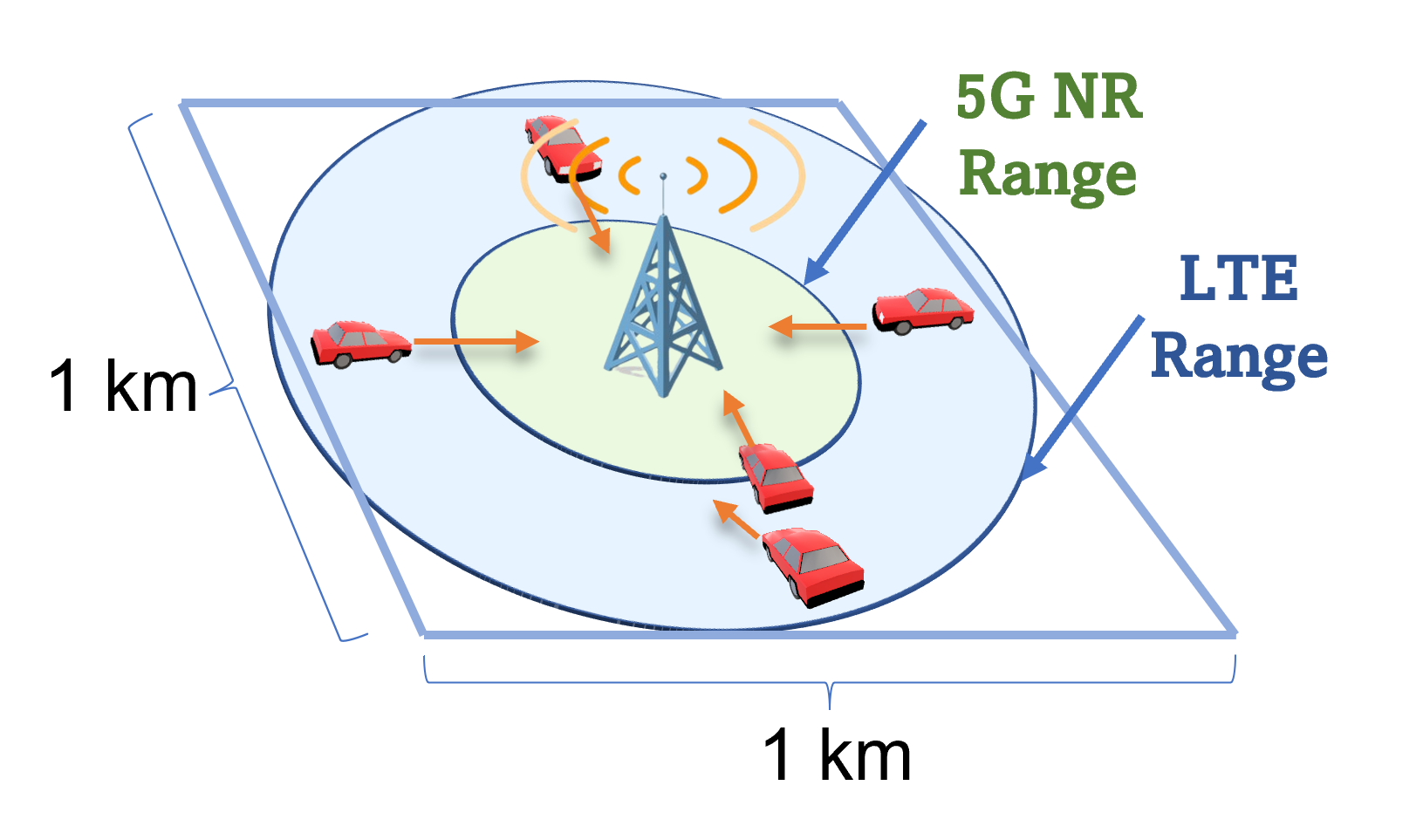}
\caption{Traffic steering simulation environment map.}
\label{map.sm}
\end{figure}

In this paper, we consider a connected urban system model with multiple users moving along roads. A multi-Radio Access Technology (multi-RAT) Base Station (BS) is set in this area, providing network coverage to users. Each user has download service requests to be satisfied by the BS, such as road condition information downloading, or other internet services like web browsing or video streaming. Note that each request has its own lifetime. These requests are made by the users and stored at the BS. A scheduler is located at the BS, serving downloading requests of all users in an efficient manner.

\subsection{Urban User Model}
Consider an urban area shown in Fig.~\ref{map.sm}, with a 1 km by 1 km map. The multi-RAT BS is located in the middle to provide better coverage. Users can be approaching or leaving the BS. Note that there can be a T-junction, crossroad, roundabout, etc.\ in the centre. In this simulation, the number of vehicles is always 5. When one vehicle leaves this area, another vehicle will be automatically generated to maintain a constant vehicle number.

\subsection{I2V Connectivity Model}
We assume that the BS provides connectivity across $R$ RATs. In the case of I2V connectivity, for each RAT, we define the downlink data rate $r_{i,j}$ achieved by the BS to vehicle $i$ over the RAT~$j$ as follows:
\begin{equation}
    r_{i,j} = \sigma_j \log_2 \left(1 + \mathrm{SINR}_{i,j}\right)
\end{equation}
where $\sigma_j$ is the bandwidth of the RAT $j$ and $\mathrm{SINR}_i$ is the Signal-to-Noise and Interference Ratio (SINR) associated with the downlink transmissions originating to vehicle $i$ over RAT $j$. In particular, we define $\mathrm{SINR}_{i,j}$ as follows:
\begin{equation}
    \mathrm{SINR}_{i,j} = \frac{G_j P_j h_j \ell^{(j)}(d_i)}{W_j + I_j}
\end{equation}
where
\begin{itemize}
    \item $G_j$ signifies the overall antenna gain
    \item $P_j$ is the transmission power for transmitting over RAT $j$
    \item $\ell^{(j)}(d_i)$ expresses the path-loss at a distance $d_i$ (between the BS and vehicle $i$) and it is defined as $C_j d_i^{-\alpha_j}$ -- $C_j$ and $\alpha_j$ are constants associated with RAT $j$
    \item $h_j$ is a random variable modelling the fast fading component and it depends on the RAT in use.
    \item $W_j$ represents the white thermal noise power. It can be seen as a constant that depends on the RAT in use.
    \item $I_j$ is the interference power. Here we assume $I_j = 0$, thus $\mathrm{SINR}_{i,j} = \mathrm{SNR}_{i,j}$.
\end{itemize}

 Considering the $i$-th vehicle $v_i$, we say that through every request, it requests a `job' ${J_i}$, which consists $T_i$ data frames, namely, $J_i = \{v_{i,1}, \ldots, v_{i,T_i}\}$) from the BS. Each data frame has an identical size, while each job is associated with a lifetime, that is, a downloading deadline. If the job has not been downloaded before its deadline, it will be discarded, thus the corresponding request is not satisfied. During transmission, if any data frame is not successfully downloaded due to any possible reasons, it would be regenerated and transmitted again.

\subsection{System Goal and tasks}
The goal of this system is to design a scheduler to dynamically meet vehicle downloading requests by using multiple RATs in an efficient manner.

\subsubsection{Caching rate}
The caching rate is the ratio of successfully transmitted bytes/packets over all requests. We define caching rate as follows: 
\begin{equation}
    CR = \frac{\text{Completed jobs}}{\text{Total requests}}
\end{equation}

\subsubsection{Latency}
Latency is calculated as follows:

\begin{equation}
    \Delta t  = \frac{t_{c}}{t_{d}},
\end{equation}
where $t_c$ is the completion time and $t_d$ is the total deadline time. Since there are two different job sizes, creating proportional latency values is fairer than calculating the remaining time in seconds.

\subsubsection{Throughput}
Throughput metric is calculated as follows:
\begin{equation}
    T  = \frac{T_{c} + T_{l}}{t},
\end{equation}
where $T_c, T_l$ are successfully transmitted bytes in completed jobs and lost jobs respectively. $t$ is the time duration in the simulation step. 

\subsubsection{Proportional Fairness}
\PZ{Fairness is a comprehensive term. Fairness in network can be based on different metrics such as latency, throughput, availability etc. In order to simplify calculations in simulation we used proportional fairness in terms of throughput distribution among users \cite{fairness2005}. } It is calculated as follows,

\begin{equation}
   F(x)  = \sum_{v} \text{log} (x_{v}),
\end{equation}
where $x_v$ is the flow assigned to the demand of vehicle $v$.

\section{Markov Decision Process Model}
The data scheduler aims to find a policy, deciding which data frame to be sent through which RAT at each time step, providing good data downloading service upon request. We assume that each vehicle is equipped with a Global Positioning System (GPS) service and the vehicle location \PZ{information are accessible by BS real-time.} 

There are two RATs available at the BS. One is the 4G (LTE), and the other is 5G NR. The communication range of LTE covers the whole simulation area, while 5G NR only covers part of the area. Simulation parameters can be found in Table~\ref{tab:env_parameter}.

\begin{table}[ht]
    \caption{Environment parameters}
    \label{tab:env_parameter}
    \centering
    \begin{tabular}{c|c|c}
    \hline
        \multicolumn{2}{c}{Parameter} & Value \\
    \hline
        \multicolumn{2}{c}{Number of vehicles} & 5 \\
    \hline
    \multicolumn{2}{c}{Vehicle speed} & 8 m/s \\
    \hline
    \multirow{2}{*}{Max communication range} & LTE & 922 m \\
    & 5G NR & 200 m \\
    \hline
        \multicolumn{2}{c}{Buffer size at the BS} & 5 \\
        \hline
        \multirow{2}{*}{Job size} & Type A & 1 MB \\
        & Type B & 10 MB\\
        \hline
        \multirow{2}{*}{Job deadline} & Type A & 0.1 s\\
        & Type B & 1 s\\
        \hline

        \multicolumn{2}{c}{Time interval between actions} & 1 ms\\
    \hline
    \end{tabular}
\end{table}

This problem can be modeled as a Markov Decision Process (MDP), with the finite set of states \eqr{\mathcal{S}}, the finite set of actions \eqr{\mathcal{A}}, the state transition probability \eqr{\mathcal{P}} and the reward \eqr{\mathcal{R}}.

\noindent\textbf{States:} At time \eqr{t}, the state ${\mathcal{S}^t}$ can be represented as
\begin{equation}
    \mathcal{S}^t = [S_U^t, S_{BS}^t].
\end{equation}
Here ${S_U^t}$ is the user(vehicle) states and ${S_{BS}^t}$ is the BS status at time ${t}$. For simplicity, we do not specify the time \eqr{t} unless necessary in the this paper. ${S_U}$ can be written as
\begin{equation}
    S_U = [(x_1, y_1, v_{x,1}, v_{y,1}), ..., (x_n, y_n, v_{x,n}, v_{y,n})]
\end{equation}
where ${x_n, y_n, v_{x,n}, v_{y,n}}$ are the geographical position and velocity of user \eqr{i}, respectively. \eqr{n} is the number of users. 

For the BS status, it contains the job buffer, status of RATs, and the link status of the `BS - vehicle' connections, represented as ${s_b}$, ${s_R}$ and ${s_c}$, respectively.
\begin{equation}
    S_{BS} = [s_b, s_R, s_c]
\end{equation}
With length of \eqr{l} the buffer status can be written as
\begin{equation}
    s_b = [(b_1, t_1, D_1), (b_2, t_2, D_2), ..., (b_l, t_l, D_l)]
\end{equation}
where ${(p_i, u_i, t_i)}$ are the number of packets left for current job \eqr{i}, the vehicle which requested job \eqr{i}, and the time left for job \eqr{i} before it would be discarded, respectively. To show the RATs status, we use binary values to describe their availability:
\begin{equation}
    s_R = [a, b], \text{s.t.} a\in[0,1], b\in[0,1]
\end{equation}
If there is no packet being downloaded through the first RAT, LTE, then \eqr{a = 1}, meaning that the first RAT is currently available, and vice versa. \eqr{b} represents the status of the second RAT, 5G NR.

As for the `BS --- vehicle' connection status, we show the potential data rate each vehicle could get through two RATs at its current position. ${s_c}$ can be written as:
\begin{equation}
    s_c = [dr_{(1,1)}, dr_{(2,1)}, dr_{(1,2)}, dr_{(2,2)}, ..., dr_{(1,n)}, dr_{(2,n)}]
\end{equation}
Here ${dr_{(m,i)}}$ means the data rate vehicle which \eqr{i} could obtain through RAT \eqr{m}.

\noindent \textbf{Actions:} The action space for the BS is defined as ${\mathcal{A} = (\{0,0\},\{0,1\},...,\{T,j\},\emptyset)}$, where ${\{T,j\}}$ means to download a packet of the ${T^{th}}$ job from the buffer to the required vehicle through RAT ${j}$, with ${\emptyset}$ meaning no action. Note that if more than one RAT is available, then the BS could choose the maximum of two actions at one time step.

\noindent \textbf{Reward:} The goal is to schedule the data downloading process so that the BS could satisfy vehicle requests in an efficient manner.

\begin{enumerate}
    \item (R1) RAT usage reward: for every unused RAT at each time step, the agent receives a reward of -1.
    \item (R2) Lost job reward: for every lost job, the agent receives a reward of -100.
    \item (R3) Successful job reward: for every successfully finished job, the agent receives a reward of +10.
    \item (R4) Latency: for every successfully finished job, the agent receives a reward of $+10(1- \Delta t$).
    \item (R5) Throughput: for all jobs, total bytes transmitted are calculated and the agent receives a reward of $+0.1T$. 
    \item (R6) Proportional fairness: for all vehicles in the environment, the agent receives a reward of $+F(x)$.
\end{enumerate}

\subsection{Heuristic action selection}

The RAT allocation simulation environment is a unique environment. Therefore, in order to compare results with the proposed approach, we designed a heuristic action selection algorithm as a baseline. The heuristic algorithm utilizes the same state information as an RL agent to decide on its actions, and then will try to provide an optimal solution according to the state. This heuristic is given in Algorithm~\ref{alg:heur1}.

\begin{algorithm}[htb]
\SetAlgoLined
\LinesNumbered
\SetKwProg{Function}{function}{}{end}
\SetKwRepeat{Do}{do}{until}
\textbf{Require:} The setting of different indoor scenarios.

\caption{Heuristic action selection}\label{alg:heur1}

    \textbf{Require:}  $s$
    
    ${Destinations} \gets{s(a:b)} $
    
    ${UE_\text{pos}} \gets{s(c:d)}$
    
    ${RAT \hskip0.5em availability} \gets{[5G, LTE]}$
    
    \If {No RAT available}{

    $Action \gets Do \hskip0.5em nothing$
    
    $Break$
    
    }
    
    \For{UEs in Buffer}{
    
    \If{$UE_x \hskip0.5em in \hskip0.5em 5G NR \hskip0.5em coverage$}{
    
    $Action \gets {[UE_x, 5G]}$
    
    $Break$
    
    \Else {
    
    $Action \gets {[UE_x, LTE]}$
    
    }
    
    }
    
    }

    \textbf{Return:}  $Action$

\end{algorithm}
This heuristic is designed to choose 5G when $UE_x$ is one of the destination UEs in the buffer. Since the 5G NR service has higher data rates, it can complete jobs faster and reduce the loss of jobs. However, in our simulation setup, we defined a 5G NR data rate as lower than LTE when a UE gets closer to the border of the 5G NR coverage area. We expected our RL agents to learn this pattern and manage resources according. 


\section{Federated Meta-learning for O-RAN Traffic Steering}
The O-RAN architecture contains RAN intelligent controllers (RIC) to control and manage radio resource allocations. Since there is more than one RIC in an NR network, we propose a distributed learning framework for AI-based resource management. 

\begin{figure}[t]
\includegraphics[width=\columnwidth]{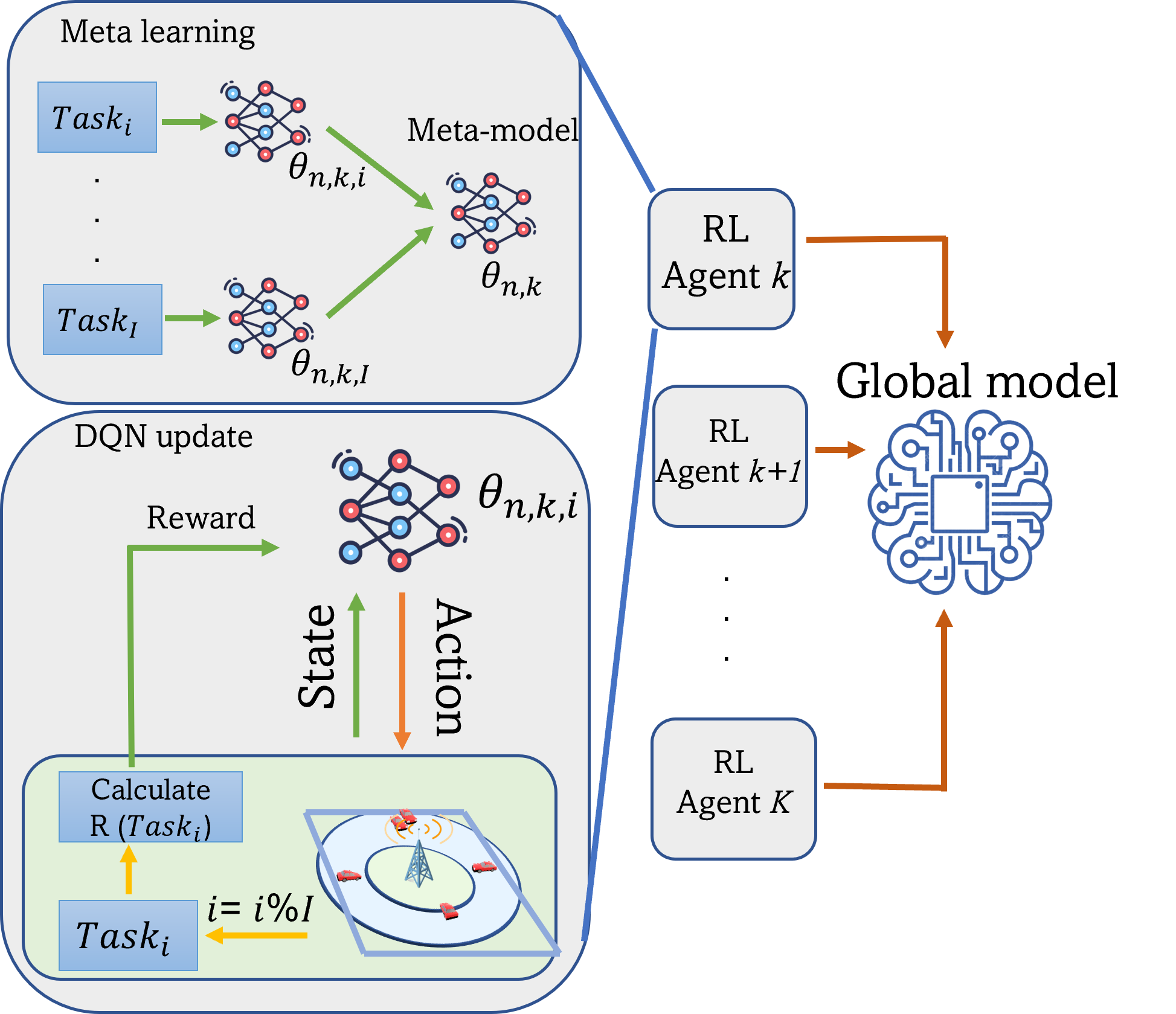}
\caption{Federated meta-learning framework set-up.}
\label{FL_setup_map.sm}
\end{figure}

\subsection{Federated Reinforcement Learning Setup}

In this paper, we use FML algorithm \cite{FL_original} to provide both a hierarchical and fast-adapting framework. FL low communication overheads permits deployment at end devices. We consider deployment at either RICs or E2 nodes \cite{ORAN_arc} as a resource management unit in the hierarchy. When the FML algorithm is deployed in the field, it will require data with the least latency. Hence, the action decisions of the RL agents will not expire for the collected state information from the environment. We assume state information of simulation is collected close to real-time. Therefore, any latency caused by transmitting data from RU to RIC is ignored in this work. The proposed FML algorithm for traffic steering in O-RAN is given in Algorithm~\ref{alg:FML}.

\begin{algorithm}[htb]
\SetAlgoLined
\LinesNumbered
\SetKwProg{Function}{function}{}{end}
\SetKwRepeat{Do}{do}{until}

    \caption{Federated meta-learning for TS simulation}\label{alg:FML}
    \textbf{Require:}
    {N, K, E, I, Num\textsubscript{vehicles}, Size\textsubscript{Buffer},$\alpha, \gamma, \beta$ }
    
    \textbf{Initialize:}
    {K environments, Vehicles, Base station, Buffer objects}
    
    \For {$n$ in $N$}{
    
    \For {$k$ in $K$}{
    
    $\theta_k \gets \theta_\text{Global}$
    
    $\text{Buffer}_k \gets Random demands$
    
    $i \gets 0$ 
    
    \For {$e$ in $E$}{
    
    $s([V_\text{pos}, V_\text{vel}, \text{Buffer}, A_{RAT}]) \gets observe env_{n,k}(t)$
    
    $a \gets \epsilon_\text{greedy}, \theta_{n,k}(s)$
    
    $R \gets step(env_{n,k}(t),a,T_i)$
    
    Store transition (s',a',R,s,a) in buffer D
    
    Sample random minibatch of transitions from D

    \If{Episode terminates}{
    
    $y \gets{R}$
    
    }
    
    \Else{
    
    $y \gets [ R + \gamma \max_{a'} Q(s',a';\theta_{n,k} )]$
    
    }
    
    Perform gradient descent step on $(y - Q(s,a;\theta_{n,k}))^2$
    
    $t \gets t + \Delta t$
    
    $i \gets i + 1$
    
    \If{$i \geq I$}{
    
    $\theta_{n,k,\text{meta}}' \gets \theta_{n,k} - \beta  \frac{1}{I}  \sum_{i=0}^{I} (\theta_{n,k} -\theta_{n,k,i}) $
    
    $\theta_{n,k} \gets \theta_{n,k,\text{meta}}'$
    
    $i \gets 0$
    
    }
  
    }
    
    Upload $\theta_{n,k}$
    
    }
    
    $\theta_\text{Global} \gets \frac{1}{K} \sum_{k=0}^{K} \theta_{n,k}$

    }
    
\end{algorithm}

In this algorithm, $N$ is the number of FL aggregation cycles, and $K$ is the number of parallel environments. Since each environment is managed by a single RL agent $K$ also equals the number of RL agents. $E$  is the number of training cycles for each RL agent, and $I$  is the number of training tasks. These tasks are used to train models in meta-learning algorithm. $\text{Num}_\text{vehicles}$ is the number of vehicles in each environment, and $\text{Size}_\text{Buffer}$ is the buffer capacity for each environment. 

After registering these data, the simulation initializes $K$ unique environments for each RL agent. These environments have different vehicle starting points in the simulation map as shown in Figure~\ref{FL_setup_map.sm}. Before beginning the training all buffers are filled with Type A and B jobs randomly as given in Table~\ref{tab:env_parameter}. Each RL agent trains its DQN model in their own environment. \PZ{DQN algorithm aims to find optimal policy to obtain optimal return according state and action. DQN algorithm estimates the action-value function by using the Bellman equation as  in equation~(\ref{eq:bellman optimality}) \cite{sutton2018reinforcement}.}

\begin{equation}
\label{eq:bellman optimality}
    Q^\pi(s,a) = \mathds{E}_{s^\prime \sim \mathcal{S}} \Big[ R + \gamma \max_{a^\prime} Q^* (s^\prime, a^\prime) \big| s,a\Big].
\end{equation}

\PZ{Here the RL agent tries to find the best action $a^\prime$ for corresponding state $s^\prime$ to find optimal policy. In order to prevent unstable updates, gradients are limited by a discount factor $\gamma $.
}
After $E$ cycles they upload their model parameters to the server. Then server aggregates these model updates with the \textit{FedAvg} algorithm \cite{FL_original} to form a global model \PZ{as given in equation~(\ref{eq:FL}). After aggregation for $K$ agents is completed, the server broadcasts global model $\theta_\text{Global}$ back to the RL agents. The RL agents use this pre-trained global model in their environment in the next FL aggregation cycle. }

\begin{equation}
    \label{eq:FL}
    \theta_\text{Global} = \frac{1}{K} \sum_{k=0}^{K} \theta_{n,k} 
\end{equation}

\subsection{Meta-learning tasks}
\PZ{The goal of meta-learning is to train a model which can rapidly adapt to a new task using only a few training steps. 
To achieve this, RL agents train the model during a meta-learning phase on a set of tasks, such that the trained model can quickly adapt to new tasks using only a small number of adaptation steps \cite{maml}. 
Since RL agents are deployed in a simulation environment where UE trajectories are likely to be unique, RL agents will try to adapt to this unseen environment. Moreover, as mentioned before besides the UE trajectories, the demands of UEs can differ in each environment. Hence, we used four different tasks to train RL agents in the meta-learning phase and observe them adapt to the fifth one. After every four rounds, RL agents update their DQN models according to equation~(\ref{eq:Meta}),}
\begin{equation}
    \label{eq:Meta}
    \theta_{n,k,\text{meta}}' = \theta_{n,k} - \beta  \frac{1}{I}  \sum_{i=0}^{I} (\theta_{n,k} -\theta_{n,k,i})
\end{equation}
\PZ{Here $\beta$ is the meta-step size, which scales model updates,  and $I$ is the number of tasks to train in a meta-learning manner. Scaling updates prevents DQN model from fully converging to a specific task. Instead of a specific task, it is expected that the DQN model converges to a point where it can adapt to an unseen task as quickly as possible.}
Six different reward functions are described in section IV for meta-learning methods. Five unique tasks are defined by using these reward functions; the tasks are listed as follows:

\begin{enumerate}
    \item Task 1 is the most comprehensive reward that an RL agent can train in this environment, which is calculated as $R_1 + R_2 + R_3 + R_4 + R_5 + R_6 $
    \item Task 2 is reward based on proportional fairness, calculated as $R_1 + R_6$
    \item Task 3 is latency-prioritized reward, calculated as, $R_1 + R_4 $ 
    \item Task 4 is throughput-based reward, calculated as $R_1 + R_5$  
    \item Task 5 is reward based on caching rate, calculated as $R_1 + R_2 + R_3$  
\end{enumerate}
Here $R$ is the reward function described in section IV. In the Reptile algorithm and the FML method, RL agents use the first four tasks for training and all methods try to adapt the 5th task in the adaptation phase, while a single DRL agent uses only Task~1 for training and tries to adapt Task~5. 

\subsection{ORAN and Traffic steering }
There are a number of resource types in the O-RAN structure and the demands of UEs can change the scheme of resource management. However, the action space for RL agents grows exponentially with number of dimensions of resources to be managed. Since most of the RL algorithms depend on explore-and-exploit methods, using multiple RL agents collaboratively likely to enhance exploration, and helps RL agents converge to better rewards faster. Hence, this is one of the major reasons for proposing the FML framework for traffic steering in O-RAN. Simulation results show even a single resource type such as the RAT allocation problem can be handled better with collaborative learning of multiple RL agents. 

\section{Simulation Results and Discussions}
\subsection{Simulation parameters}
In traffic steering simulations, we observed several parameters in the simulation to see how FL performs under different conditions. In most of the simulation runs, FML performed significantly better than other methods, but in some occasions heuristic and FML performance was almost equal. We ran a simulation with 20 different parameter combinations, and an average performance comparison is achieved with the parameters as, the number of FL aggregating cycles is $N = 10$, the number of training tasks is $I = 4$, number of created environments (and RL agents) in FL network is $K=5$, number of episodes before aggregation is $E = 100$ and time interval between environment steps is $\Delta t = 1$ms.

Note that, in order to have a fair comparison, single DRL and Reptile-based DRL agents are trained equally as much as RL agents in the FML framework. In this case, it is $N * E$. 
 
\subsection{Caching performance results}
In the simulation, we tracked lost packets and lost bytes in each lost packet to calculate a caching rate performance indicator. Each method has been run for 10 validation runs in an unseen environment (unique environments in terms of UE trajectory) in the training phase. After 10 runs for each parameter combination, we averaged the caching results to get the final score for each approach which is given in Table \ref{tab:performance}.
\begin{table}[h]
    \caption{Caching-rate performance comparison}
    \label{tab:performance}
    \centering
\begin{tabular}{lrr} 
\hline
Method       & Packets & Bytes \\ \hline
Single RL    & 81\%   & 90\%    \\ \hline
Heuristic    & 78\%   & 88\%    \\ \hline
Reptile      & 86\%   & 92\%    \\ \hline
\textbf{Proposed FML} & \textbf{89}\%    & \textbf{95}\% \\
\hline
\end{tabular}
\end{table}

\PZ{As shown in Table~\ref{tab:performance}, the proposed FML approach achieves the highest caching rate performance amongst the compared methods in terms of packets and bytes. We calculated these results as the ratio of successfully transmitted bytes/packets over total requests.}
\subsection{Adaptation performance}
\begin{figure}[h]
\includegraphics[width=\columnwidth]{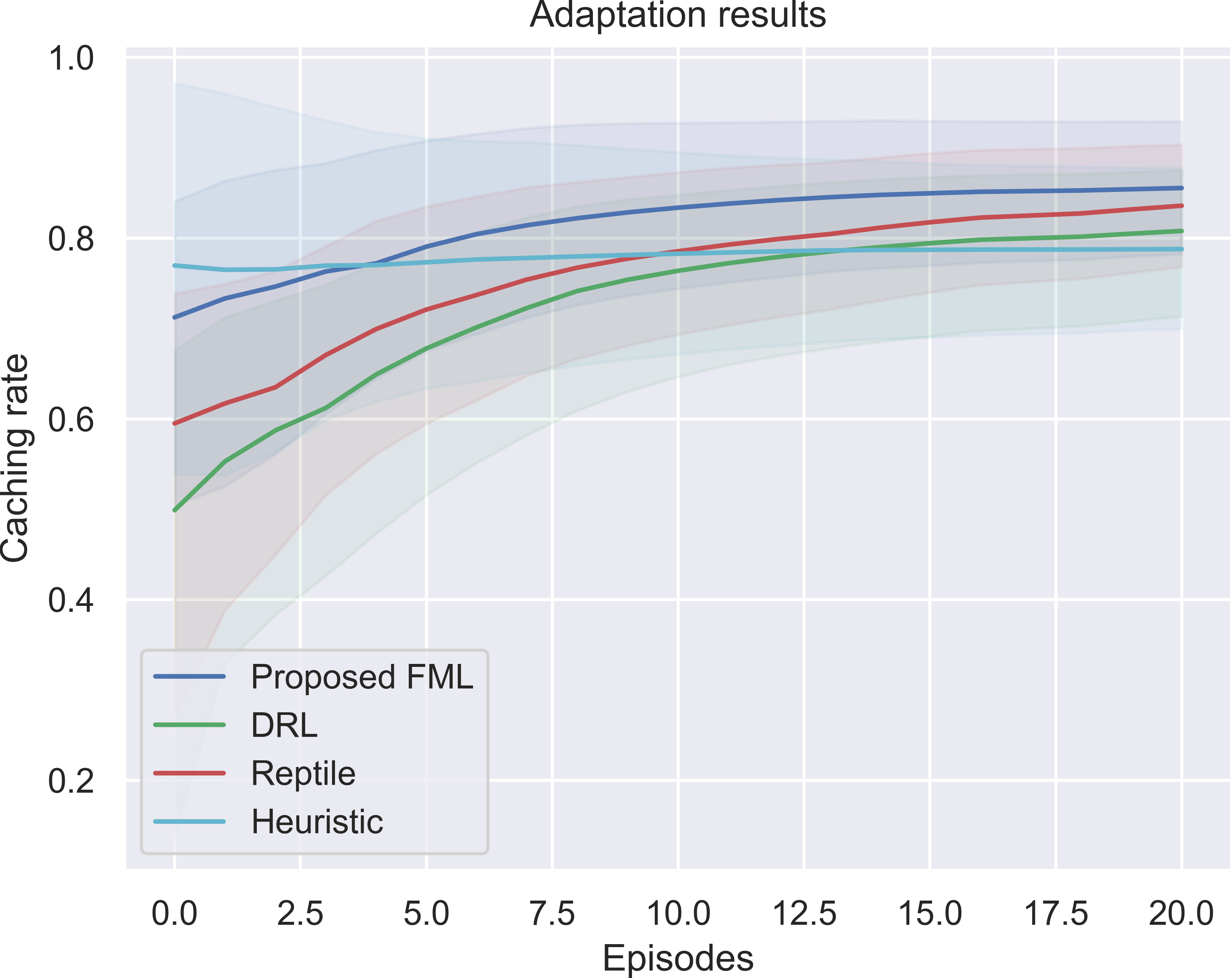}
\caption{Adaptation performance comparison to unseen environments.}
\label{Adapting.sm}
\end{figure}

Simulation results show that FML method improves adaptation performance from the very first training episodes. As a quantitative comparison, RL-based algorithms try to reach heuristic method performance as soon as possible. According to Figure~\ref{Adapting.sm}, the FML algorithm achieves heuristic equivalent performance the quickest among all methods. Zero-shot adaptation performances of different methods are compared in Table~\ref{tab:adapt_perf}, where HEP is heuristic equivalent performance.

\begin{table}[h]
    \caption{Adaptation performance comparison}
    \label{tab:adapt_perf}
    \centering
\begin{tabular}{lrrr} 
Method    & 0-Shot (mean) & 0-Shot (std) & Episode of HEP \\ \hline
Single RL & 51\%         & 13\%        & 13                                   \\ \hline
Reptile   & 60\%          & 12\%         & 9                                    \\ \hline
\textbf{Proposed FML}       & \textbf{72\%}          & \textbf{10\%}         & \textbf{3}   \\ \hline
\end{tabular}
\end{table}

%

\subsection{Discussions}

As shown in Table~\ref{tab:performance}, FML has demonstrated to be a better solution for designed traffic steering simulation. Even though we observed similar training performances for a single RL agent and FML agent, FML performed better in an unseen environment. The FL framework takes advantage of collecting information from various environments, and so it becomes easier to adapt to a new environment. There are cases where a single learner performs better than the global model because of the issue of collected not independent and identically distributed (non-iid) data. Nevertheless, there are other solutions to prevent performance deterioration at the global model \cite{noniid2}.

In future work, we will add new resource types to this environment. Traffic steering is a comprehensive use case, since the 5G NR standard allows service providers to collect various communication-related data from UEs, it is more likely to have AI/ML-based solutions for such problems \cite{5G_survey}.

\section{Conclusion}

In this paper, we have focused on the generalization of different tasks and environments by using the FML framework. As a use case, we used traffic steering in O-RAN architecture. We designed a traffic steering environment to investigate how DRL-based learning algorithms perform in such an environment. While unique environments are created for every RL agent in a stochastic way, RL agents try to manage RIC and allocate RAT services among UEs in the environment. We have analyzed the convergence speed of the DRL algorithm that uses a single task and single environment. Another DRL algorithm that uses multiple tasks in the Reptile algorithm and single environment and proposed FML framework that uses multiple tasks and multiple environments to train RL agents. Simulation results 
confirm the effectiveness of our proposed algorithm. Future work can investigate how the FML framework deals with managing multiple radio resources on RICs.

\section*{Acknowledgment}
This work was developed within the Innovate UK/CELTIC-NEXT European collaborative project on AIMM (AI-enabled Massive MIMO).
\bibliographystyle{IEEEtran} %
\bibliography{IEEEabrv,references} 

\end{document}